\documentclass[%
 reprint,
superscriptaddress,
 amsmath,amssymb,
 aps,
]{revtex4-2}

\usepackage[english]{babel}

\usepackage{amsmath}
\usepackage{graphicx}
\usepackage[autostyle]{csquotes}
\usepackage{multirow}
\usepackage{mathrsfs}
\usepackage{upgreek}
\usepackage{lineno}
\usepackage{float}
\usepackage{xcolor}

\begin{document}

\title{Scatterless interference:
\\
Delay of laminar-to-turbulent flow transition by a lattice of subsurface phonons}

\author{Mahmoud I. Hussein}
\email[]{mih@colorado.edu}
\affiliation{Smead Department of Aerospace Engineering Sciences, University of Colorado, Boulder, Colorado 80303}
\affiliation{Department of Physics, University of Colorado, Boulder, Colorado 80302}
\author{David Roca}
\affiliation{Centre Internacional de M\`{e}todes Num\`{e}rics en Enginyeria (CIMNE), Barcelona 08034, Spain}
\affiliation{Universitat Polit\`{e}cnica de Catalunya, ESEIAAT Campus Terrassa UPC, Terrassa 08222, Spain}
\author{Adam R. Harris}
\affiliation{Smead Department of Aerospace Engineering Sciences, University of Colorado, Boulder, Colorado 80303}
\affiliation{Materials Science and Engineering Program, University of Colorado, Boulder, Colorado 80303}
\author{Armin Kianfar}
\affiliation{Smead Department of Aerospace Engineering Sciences, University of Colorado, Boulder, Colorado 80303}

\begin{abstract}
Wave interference has historically relied on scattering objects placed within the wave domain. Here, we introduce a fundamentally new mechanism: scatterless interference induced by a lattice of subsurface phonon motion beneath a smooth wall interfacing with an unstable laminal channel or boundary-layer flow. The subsurface consists of a wall-parallel lattice of wall-normal frequency-dependent phononic structural units, each designed to locally respond to a growing flow perturbation in an out-of-phase manner, dynamically influencing it at the point of interaction. Collectively, the lattice induces an interference effect that causes the kinetic energy of the flow instability to decay downstream, thereby delaying laminar-to-turbulent transition. To guide the design of the phononic subsurface lattice, a Bloch-wave unit-cell analysis is developed for flow perturbations, and direct numerical simulations validate the concept.~This work establishes scatterless interference as a distinct physical phenomenon and represents a paradigm shift in the design of aerodynamic and hydrodynamic surfaces—moving beyond streamlined shaping to leveraging subsurface phonon engineering for drag reduction and enhanced performance. 
\end{abstract}


\maketitle

\noindent \textbf{Introduction} \\
The control of wave interference through the deliberate placement of scattering objects has been a cornerstone principle in wave physics for over two centuries.~Its classical origins  trace back to Thomas Young’s double-slit experiment~\cite{young1804bakerian}, which first demonstrated the wave nature of light via interference between scattered paths. This concept evolved into a general principle applied across multiple disciplines.~In electromagnetics, the development of photonic crystals~\cite{yablonovitch1987inhibited,john1987strong} relied on arrays of dielectric scatterers directly interacting with the wave field to trigger spatial constructive and destructive interferences that bring rise to Bragg scattering.~A similar mechanism emerged in acoustics and elasticity, where periodic voids or inclusions, forming phononic crystals~\cite{sigalas1992elastic,kushwaha1993acoustic}, have enabled the formation of wave interference patterns and spectral band gaps through in-domain scattering.~Even in quantum mechanics, interference effects typically arise through the interaction of wavefunctions with potentials or obstacles embedded in the same domain of motion. These classical and modern mechanisms share a foundational attribute: the structures responsible for interference are located within the same spatial domain as the wave field they manipulate~\footnote{The only exception is having active external excitations to trigger the scattering~\cite{vasquez2011exterior,russomanno2012periodic, makris2017wave}}.~To date, this paradigm has remained essentially unchallenged across physics.~In this work, we introduce a fundamentally new mechanism of passive wave interference that breaks from the conventional framework: \textit{scatterless interference} induced by a lattice of phononic subsurface units. In sharp contrast to prevalent scattering mechanisms, the interference patterns in our fully passive system arise not from scatterers embedded within the wave domain, but from elastic phononic structures buried beneath the surface—outside the spatial domain occupied by the propagating waves. We present this new physical mechanism in the context of the long-standing classical fluid dynamics problem of laminar-to-turbulent transition.\\
\noindent \textbf{\textit{Fluid-structure interaction}}\\
\indent The laminar-to-turbulent transition in wall-bounded flows has been the subject of extensive research, resulting in a rich build-up of knowledge on how transition unfolds across different flow regimes~\cite{morkovin1969many}.~A ``natural” transition pathway typically involves the growth of infinitesimal flow perturbations$-$also referred to as disturbances~\footnote{These terms are used interchangeably in the literature.}$-$which represent inherent unstable modes within the flow.~Without intervention, these modes may advance to nonlinear amplification, leading to flow structure breakdown and the ultimate evolution into fully developed turbulence~\cite{morkovin1969many}.~A dominant primary mode in air or water channel flows, as well as boundary layers, is known as the Tollmien-Schlichting (T-S) wave~\cite{Tollmien_1928,Schlichting_1933}. Unstable T-S waves take a vorticial form and travel with the mean flow, growing over an identifiable narrow band of frequencies that can be predicted with linear stability analysis~\cite{schubauer1947laminar,mack1984boundary}.~While T-S waves are not always the direct cause of transition, especially under complex or realistic operating conditions, they have been the focus of extensive research~\cite{carpenter1985hydrodynamic}.~This is due to their fundamental nature, and because they serve as a platform for development of new technologies for laminar flow control involving more complex types of instabilities. \\
\begin{figure*} [t]
\centering
\includegraphics[width=1\textwidth]{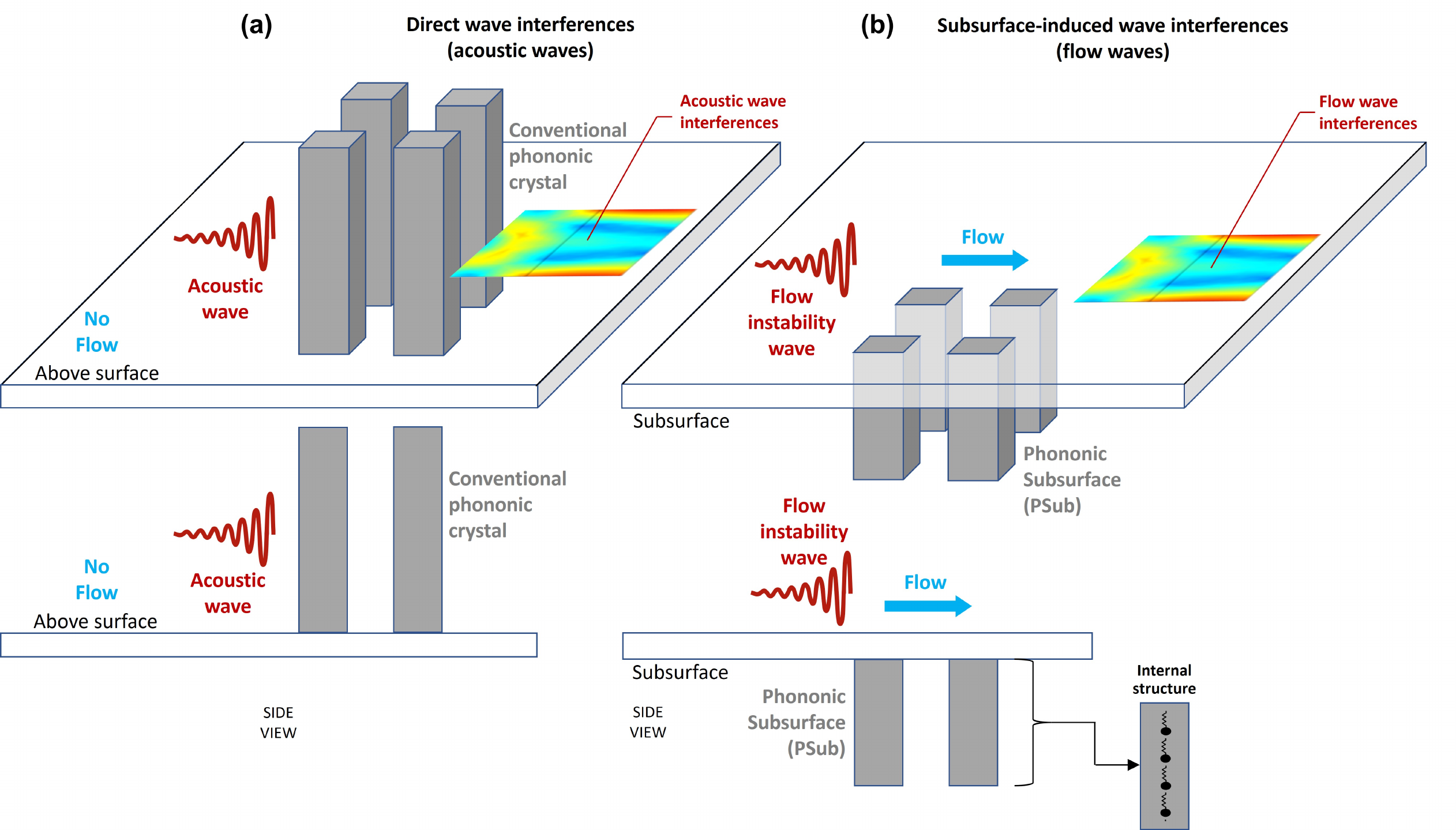}
\caption{\textbf{Illustration of scatterless interferences:} (a) Classical scattering of acoustic waves by a lattice of elastic objects, where the interfering waves and the objects are in the same space. (b) Scatterless interferences of flow perturbations waves by a lattice of PSubs, where the waves and the PSubs are not in the same space.}
\label{fig:F1}
\end{figure*}
Aside from the conventional practice of streamline shaping of surfaces, numerous research approaches have been proposed in the last couple of decades aimed at passively mitigating the undesirable effects of fluid-structure interactions.~These include installation of riblets over the surface~\cite{Walsh_1978,Garcia_2011}, creation of artificial surface roughness~\cite{Cossu_2002,Fransson_2005}, drilling of holes to form a porous surface~\cite{fedorov2001stabilization,Abderrahaman_2017}, or coating the surface with a compliant (low-stiffness) viscoelestic material~\cite{Kramer_1957,Benjamin_1960,Bushnell_1977,Gad-el-Hak_1984,Carpenter_1985,Lucy_1995,Davies_1997}.~These approaches, however, are not based on a synchronization with the frequency, phase, and wavevector characteristics of the flow instabilities, and are therefore limited in their effectiveness.~An ideal intervention requires a tailored solution, with mechanistic precision, to create a passive and responsive control stimulus that accounts for the dynamical properties of the underlying flow transition mechanisms.~In 2015, the general concept of flow control by~\textit{subsurface phonons} was introduced as an approach capable of achieving this level of precise wave-synchronized control of flow instabilities~\cite{Hussein_2015}.~A phononic subsurface (PSub) comprises a synthetically designed architected material affixed beneath~\footnote{Installation of a PSub may effectively be concealed because it is placed underneath the surface. In other words, an external observer viewing a PSub-installed airplane wing would view a normally looking surface and would not be able to identify if a PSub is present or not.} the surface exposed to the flow (e.g., of a wing or vehicle body). The function of the PSub is to manipulate small-amplitude vibrations on the surface, and by extension the flow perturbations near the wall that are responsible for transition. PSubs may be designed to passively respond to the flow instabilities in an out-of-phase manner, creating stabilization, or in an in-phase manner, creating destabilization$-$either function is realized \textit{a priori} by design, or in the future may be switched or tuned in real time by actuation.~In past computational investigations, a PSub has been applied as a solitary unit~\cite{Hussein_2015,Barnes_2021,kianfar2023phononicNJP,willey2024tollmien,schmidt2025perturbation} or as a contiguous layout of units distributed along the streamwise direction~\cite{michelis2023attenuation,kianfar2023local}.  \\
\indent Despite the rapid progress of research on flow control by subsurface phonons,  critical aspects remain to be addressed for the general concept to reach its potential. Two key limitations of previous demonstrations are that the PSub effect is not effective downstream to the PSub location, and that it is applicable only to unidirectional instability waves.~The former must be resolved to enable transition delay, and the latter  is significant for real-world flow control$-$where cross-flow scenarios may be encountered, for example.~While the downstream control objective was addressed with a ``multiple-input-multiple-output” PSub configuration, offering a remarkable explicit display of transition delay~\cite{Barnes_2021}, that approach is inherently limited to unidirectional instabilities because it dictates that a single PSub must interact with more than one flow point requiring a phased relation to be tuned specifically along a certain direction.~Furthermore, the fixed distance between the input/output points limits the approach to a narrow band of perturbation wavenumbers. Schmidt et al.~\cite{schmidt2025perturbation} explored the concept from an alternative angle: rather than engage with subsurface phonons to inhibit the perturbation production mechanisms (as done in all previous PSub studies), they designed their PSub based on pass-band motion to absorb and trap, by filtering, the energy of undesirable fluctuations. Actively controlled time-varying material properties were used for demonstration, however the principle may be implemented passively using nonlinear effects.~This approach, which is applicable to only the function of stabilization, permanently attracts the energy that is continuously being transferred from the mean flow to the instabilities, instead of reducing it at the source~\cite{kianfar2023phononicNJP}.~This energy will accumulate in the subsurface structure requiring its ultimate dissipation as heat. \\
\noindent \textbf{\textit{Scatterless interference}}\\
\indent In this work, we present the concept of a \textit{lattice of PSubs} comprising a collection of individual PSub units laid out following a square or hexagonal lattice symmetry~\footnote{Other lattice symmetries may also be considered in the future.}.~Similar to acoustic or elastic waves propagating around or through a lattice of rigid or elastic scatterers, respectively (see Fig.~\ref{fig:F1}a), we design our configuration on the basis of a rigorous Bloch wave analysis where the field variable is the flow perturbations.~Yet, in our system there are no scatterers. In contrast to classical scattering problems where the interfering waves and the scattering objects are located in the same space, the flow instability waves interfere as a result of the local influence of each PSub on the perturbation velocity components. Here we recall that the PSubs are located beneath the spatial domain of the flow (see Fig.~\ref{fig:F1}b). After developing Bloch's theorem for this unique problem, we demonstrate by direct numerical simulation (DNS) of the coupled fluid-PSub lattice systems sustained downstream reduction in the perturbation kinetic energy (KE), which indicates a delay in laminar-to-turbulent transition. Moreover, as dictated by the PSub lattice symmetry, this approach is immune to any changes in the direction of propagation of the instability waves and may be tuned to accommodate a relatively wide range of frequencies and wavenumbers along each direction$-$similar to the classical acoustic Bragg scattering problem which we also demonstrate to provide a direct analogy and comparison.  \\
\begin{figure}
\includegraphics[width=1\columnwidth]{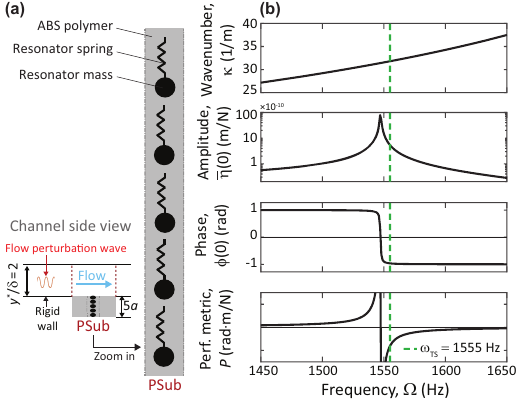}
\caption{\textbf{PSub design and its dispersion and vibration response characteristics:} (a) Schematic of locally-resonant elastic metamaterial-based PSub unit utilized in this study. The length of the unit cell is $a_{\rm Psub}=1$ cm yielding a total PSub length of 5 cm. Each PSub is installed in the flow subsurface with its top face directly exposed to the flow. Flow instabilities, e.g. T-S waves, will continuously excite the PSub at the top edge and the PSub, in turn, will respond at the same point. Depending on the frequency, the PSub will respond according to its dynamical characteristics. Following an out-of-phase or in-phase response, the PSub motion will respectively impede or enhance the energy extraction from the mean flow into the perturbation field~\cite{kianfar2023phononicNJP}.~This passive process will repeat and cause sustained control of incoming instability waves near the wall at the location of the PSub. (b) Four key plots that characterize the PSubs dynamics. From top to bottom, the following are shown: dispersion curves for PSub unit cell, steady-state vibration amplitude and phase response of the PSub top edge when excited at the same location, and performance metric obtained by multiplying the amplitude by the phase. All plots are obtained by analyzing a stand-alone finite-element model of the PSub.}
\label{fig:F2}
\end{figure}
\hspace{1cm}\\
\noindent \textbf{Results} \\
\indent We form our lattice of PSubs using identical PSub units, all installed in a channel model. We select a channel flow problem for simplicity, but all the underlying concepts are readily applicable to boundary layer flows. The design of the PSub unit used in this investigation is shown in Fig.~\ref{fig:F2}a. It consists of an elastic rod, composed of five unit cells, with each unit cell accommodating a local resonator (which in practice may be realized as a cantilevered beam or pillar).~The dispersion curves and the amplitude and phase frequency-response characteristics for this nominal PSub configuration is shown in Fig.~\ref{fig:F2}b. The product of the amplitude and phase yields a performance metric~\cite{Hussein_2015} which provides an \textit{a priori} prediction of the behavior of the PSub once passively engaged with the flow$-$this quantity is also plotted in Fig.~\ref{fig:F2}b.\\
\begin{figure*} 
\centering
\includegraphics[width=1\textwidth]{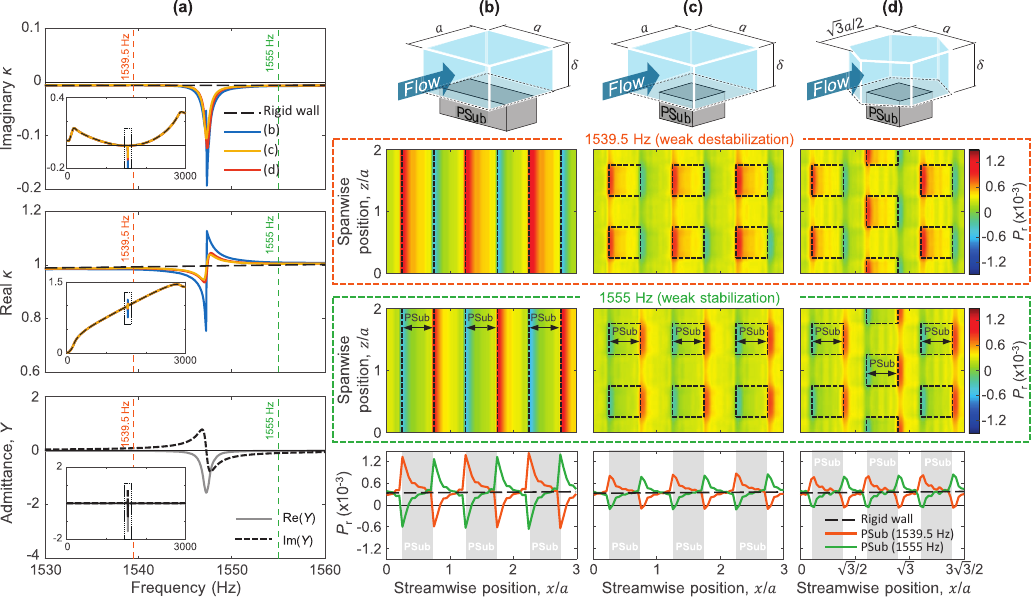}
\caption{(a) \textbf{Bloch analysus or flow perturbations: Dispersion curves for selected unstable flow mode in the presence of a PSub lattice}. The first and second rows correspond to the imaginary and real components of the wavenumber $\kappa$ for the rigid-wall case and the PSub cases in the full-span, square-lattice, and hexagonal-lattice configurations [depicted by the (b), (c) and (d) unit cell schematics, respectively]. The third row shows the real and imaginary components of the PSub's admittance $Y$ at different frequencies. The dashed vertical lines indicate the frequencies for which the average modal perturbation KE production rate, $P_\text{r}$, distributions on the right plots have been obtained. Results for each frequency are depicted in their corresponding panel. The first panel (in orange) shows a weak destabilizing case (at a frequency of 1539.5~Hz) and the second panel (in light green) is for a weak stabilizing case (at a frequency of 1555~Hz). The curves below the panels correspond to the averaged production rate evolution obtained for both the weak destabilizing and weak stabilizing cases compared to that of the rigid case. As a reference, the average has been evaluated in the vicinity of the PSub wall (in the range of 10\% of half of the channel's height), and all the modal amplitudes have been normalized to make them coincide with that of the rigid-wall case at the left edge of the first unit cell.}
\label{fig:F3}
\end{figure*}
\begin{figure*} 
\centering
\includegraphics[width=1\textwidth]{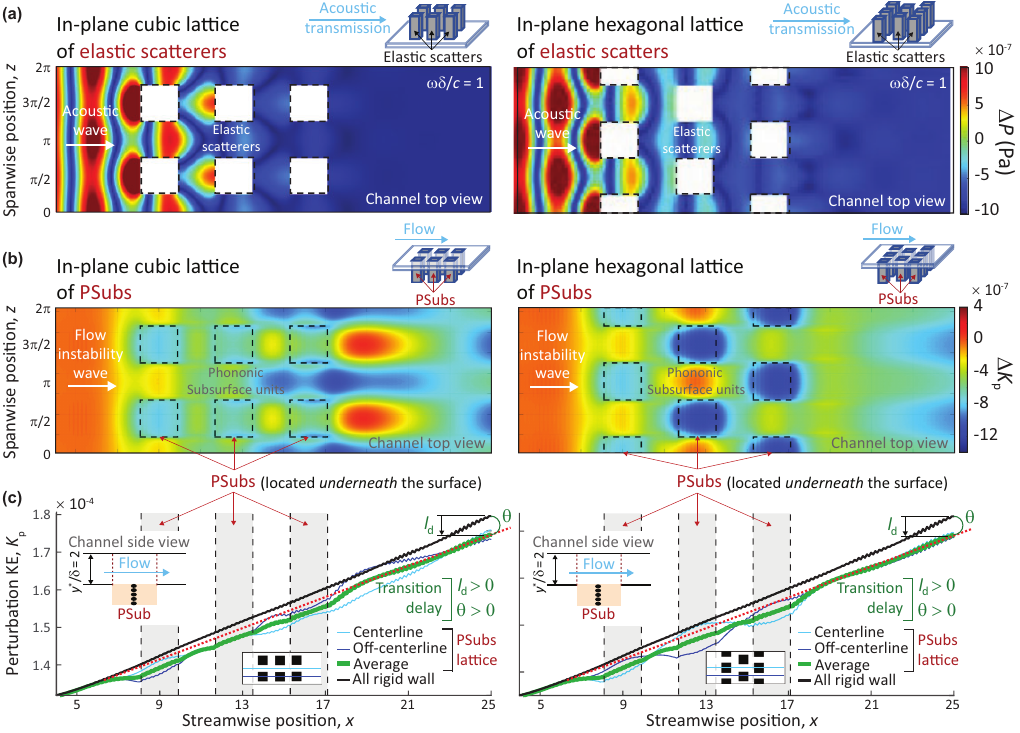}
\caption{\textbf{Demonstration of scatterless interferences of flow perturbations by DNS:} (a) Acoustic wave propagation in a space covered by a lattice of elastic scatterers, considering square symmetry (left) and hexagonal symmetry (right). (b) Time-averaged flow perturbation KE as a function of the streamwise and spanwise directions for flow in a channel with PSubs laid out following both symmetries. (c) Time- and spanwise-averaged perturbation KE as a function of the streamwise position. The results in (b) and (c) indicate interference patterns, even though the surface is flat and smooth and there are no objects in the flow space. These interferences stem for the collective action of the local effect by each PSub. Figure 4(c) shows time-averaged perturbation KE on and off the centerline of the channel, reflecting the effects of the different symmetries of the square and hexagonal lattices. The time- and spanwise-averaged perturbation KE is shown in dark green, where the angle $\theta$ at the end of the channel indicates a decrease in the slope of $K_{\text{p}}$ compared to the rigid-wall case, and hence transition delay downstream of the PSub lattice. The angle is measured as $\theta=4.4^{\circ}$ and $\theta=4.6^{\circ}$ for the square and hexagonal cases, respectively, indicating the transition is delayed further by the hexagonal PSubs lattice.}
\label{fig:F4}
\end{figure*}
\indent Figure~\ref{fig:F3} shows the results of applying Bloch analysis over a unit cell comprising the flow interacting with the PSub in three configurations, a full-span PSub, and the square and hexagonal lattice arrangements [see schematics (b), (c) and (d) in Fig.~\ref{fig:F3}]. The PSub interaction with the flow has been accounted for through a frequency-dependent complex-valued admittance, denoted by $Y$ in Fig.~\ref{fig:F3}(a). The admittance is obtained, for a given frequency, from the solution of the PSub system of equations~\cite{Barnes_2021} and it is closely related to the performance metric. We use this function alongside transpiration boundary conditions to find the most unstable eigenvalue$-$and associated eigenmode$-$from the standard Orr-Sommerfeld equations modified to accommodate the periodicity imposed by the application of Bloch's theorem in terms of the T-S wave's wavenumber, which corresponds to the eigenvalue. The mathematical details of the formulation and definitions of key quantities can be found in the Supplementary Material.~As a result of this analysis, the dispersion curves corresponding to the unstable flow mode in Fig.~\ref{fig:F3}(a) have been obtained.~These results are complemented by computing the average perturbation kinetic energy, denoted by $K_{\text{p}}$, and the rate of production of that energy, denoted by $P_{\text{r}}$.~The perturbation KE has been computed from the eigenmodes associated with the unstable eigenvalues, normalized by the corresponding value for the rigid case. The $P_{\text{r}}$ quantity, as obtained from the Bloch analysis, is plotted in Fig.~\ref{fig:F3}b-c. The reader is referred to Supplementary Materials for calculation details. Results for both a weak destabilizing case and a weak stabilizing case, at frequencies of 1539.5~Hz and 1555~Hz respectively, have been averaged over a flow domain close to the PSub wall, in particular along 10\% of the channel's half height. Contour plots showing the $P_\text{r}$ spanwise distribution over three unit cells (in the streamwise direction) are given in Fig.~\ref{fig:F3}(b) to (d) for the full-span and for each of the lattice configurations. The average $P_\text{r}$ values over a spanwise period are shown in the last row of Fig.~\ref{fig:F3}(b) to (d) to summarize the results. The different cases are compared with the results for the rigid case, to show the spatial stabilizing/destabilizing effects when the flow interacts with each PSub unit.\\
\indent The ultimate test for any flow control approach is through high-fidelity simulations, particularly DNS of the Navier-Stokes equations, as these resolve the high spatial and temporal frequencies of the flow perturbations$-$which are central to the transition mechanism. We execute simulations in a three-dimensional (3D) computational domain and retain all nonlinear terms. In Fig.~\ref{fig:F4} we present DNS results for the two lattice configurations, and also show the results of an analogous acoustic scattering simulation for comparison. The coupled fluid-structure simulations were run at $Re=7500$ with a 1555~Hz unstable T-S wave input at the left end of the channel. See the Models and Methods sections for more details on the model, simulation parameters, and numerical techniques used. The results in Figs.~\ref{fig:F4}b and~\ref{fig:F4}c show a sustained reduction in $K_{\text{p}}$ downstream of the PSub lattice, with a slightly stronger effect by the hexagonal arrangement compared to the square arrangement. This behavior is indicative of a delay in laminar-to-turbulent transition downstream of the PSub lattice region.\\

\noindent \textbf{Discussion} \\
\indent The performance metric for the designed PSub unit shown in Fig.~\ref{fig:F2}b shows that the unstable wave frequency of 1555~Hz falls at a negative value, indicative of a stabilization effect. While the surrounding negative region is relatively narrow$-$since it is associated with a given structural resonance of the finite PSub structure, future work will explore strategies to broaden the frequency range of operation. \\
\indent Results from the Bloch analysis show a correlation between the dispersion curves (in Fig.~\ref{fig:F3}) and the PSub performance metric (in Fig.~\ref{fig:F2}).~It is observed that the intensity of the T-S wave's instability$-$in this case given by the imaginary wavenumber$-$rapidly grows in the vicinity of the resonance frequency, which helps enhance the stabilizing/destabilizing effects. Furthermore, the rate of production $P_{\text{r}}$ within the PSub region consistently increases in the destabilizing case and decreases$-$even reaching negative values$-$in the stabilizing case, in agreement with the stand-alone performance metric predictions (see Fig.~\ref{fig:F3}(b) to (d)). Interestingly, $P_\text{r}$ also exhibits an opposite effect at the trailing edge of each PSub unit, which is in accordance with what we observe in the fluid-structure interaction (FSI) simulations (see, for instance, how the $K_{\text{p}}$ slightly increases in Fig.~\ref{fig:F4}(b), for both lattice configurations, each time the flow ``leaves" a PSub unit behind despite being stabilizing cases). This analysis also shows that the leading and trailing edge changes in $P_\text{r}$ are stronger in the full-span case compared to the two lattice configurations, except that this case does not experience spanwise interactions. While both square and hexagonal PSub lattices are effective in sustaining downstream stabilization, both the Bloch analysis and simulations indicate that the hexagonal configuration is slightly superior$-$its $K_{\text{p}}$ slope in Fig.~\ref{fig:F4}c is $\theta=4.6^{\circ}$ compared to $\theta=4.4^{\circ}$ for the square lattice. Upon further design optimization of both the PSub unit and the lattice configuration, stronger downstream reductions in $K_{\text{p}}$ are attainable. It is intriguingly observed that the flow perturbations ``scatterless" interferences resemble the interferences triggered by acoustic scattering, for both lattice configurations.  \\
\indent In conclusion, the mechanism of scatterless interference reported here reveals a unique class of wave behavior that to our knowledge has not been previously reported in any branch of physics.~The concept comprises subsurface phononic units arranged with full spatial flexibility and designed to vibrate at specific frequencies, enabling precise, tunable, and frequency-selective interference patterns in the overlying wave field. The physical platform we employ to demonstrate this mechanism is an unstable laminar flow field subject to perturbation waves—a canonical setting for studying aerodynamic or hydrodynamic instability.~We have shown that by engineering the phonon properties of the subsurface$-$virtually entirely concealed beneath the flow-exposed surface$-$it is possible to exert full command over the very nature of the underlying fluid-structure interaction.~This passive intervention offers precise mechanistic and tunable control of the behavior of wall-bounded flows in general.~Furthermore, our findings demonstrate that the principles of Bloch wave analysis, traditionally applied to electronic, photonic, acoustic, and elastic waves~\cite{Bloch1929,hussein2009reduced}, can be effectively extended to flow perturbations influenced by PSubs. By leveraging locally resonant metamaterial-based individual PSub units—each acting in the wall-normal direction and collectively working in tandem as a subsurface lattice along the wall-parallel plan—we have demonstrated  that flow instabilities can be manipulated through interference effects in a manner analogous to wave phenomena in periodic acoustic or elastic media, and more broadly for elecrtons and photons.~This approach enables passive control over critical processes such as the transition to turbulence, with the PSub lattice facilitating attenuation of unstable T-S waves downstream, while being robust to variations in wavenumber and direction of propagation.~Other types of flow instabilities may be similarly treated in future studies.~With the ongoing influx of new concepts from phonon engineering and metamaterials~\cite{hussein2014dynamics,Jin_2021,jin20252024}, the prospects of further future improvements of the performance following this framework is highly accessible, and so is the path to practical implementation given the rapid advancements in fabrication technology~\cite{steijvers2023incorporating,wang2025industrial}.~A wide range of flow-control applications stand to benefit, spanning from hypersonic airflows to low-speed liquid transport. It is conceivable that within this framework future aircrafts will maintain a fully laminar boundary layer over their surfaces, enabling transformative improvements in fuel efficiency.~In addition, PSub lattices may be simultaneously designed to inhibit aeroacoustic emissions, offering the additional rewards of improved passenger comfort and reduced environmental noise pollution.~Similar advances will materialize for hydroacoustic emissions as well.~Lastly, extensions of the scatterless interference principle to other branches of wave physics such as quantum dynamics and electromagnetics may be explored.\\

\noindent \textbf{Models and Methods} \\
\indent Governed by the 3D Naiver-Stokes equations, a series of direct numerical simulations (DNS) are run for incompressible channel flows.~The velocity vector solution is expressed as ${\bf u}(x,y,z,t)=(u,v,w)$ with components in the streamwise $x$, wall-normal $y$, and the spanwise $z$ directions, respectively, where $t$ denotes time. The DNS is run for a Reynolds number of $Re=\rho_\mathrm{f} U_{\rm c}\delta/\mu_\mathrm{f}=7500$ based on a centerline velocity $U_{\rm c}=17.12$ $\mathrm{m/s}$ and a half-height of the channel $\delta=4.38\times10^{-4}$ $\mathrm{m}$. Liquid water is considered with a density of $\rho_\mathrm{f}=1000$  $\mathrm{kg/m^3}$ and dynamic viscosity of $\mu_\mathrm{f}=1\times10^{-3}$ $\mathrm{kg/ms}$. All subsequent quantities, unless mentioned explicitly, are normalized by the channel's velocity $U_{\rm c}$ and length $\delta$ scales. The channel size is $0 \leq x \leq 30$, $0 \leq y \leq 2$, and $0 \leq z \leq 2 \pi$ for the streamwise, wall-normal, and spanwise directions, respectively.~At the inlet of the channel, a fully developed Poiseuille flow is superimposed with an unstable T-S mode obtained from linear hydrodynamic stability analysis governed by the Orr-Sommerfeld equation \cite{Orr1907,Sommerfeld1909} and solved for the same $Re$. A strongly growing eigensolution is selected that has complex wavenumber $\alpha=1.0004-\mathrm{i}0.006171$ and real non-dimensional frequency $\omega_\mathrm{TS}=0.250$. Following dimensional analysis, the frequency of the corresponding T-S wave is $\Omega_\mathrm{TS}={\omega_\mathrm{TS}}{U_{\rm c}}/{2\pi}{\delta} = 1555$ $\mathrm{Hz}$. To ensure outgoing wave motion on the other side of the channel, the disturbances are smoothly brought to zero by attaching a non-reflective buffer region at the outlet~\cite{Dana91,Saiki93,Kucala14}.~Periodic boundary conditions are applied in the spanwise direction.~At the top and bottom walls no-slip, no-penetration boundary conditions are applied, except within the control region in the streamwise direction where the rigid wall is replaced by a lattice of PSubs at the bottom wall. Within each PSub control region, the fluid-structure coupling is enforced by means of transpiration boundary conditions~\cite{Lighthill_1958,Sankar_1981,Hussein_2015,kianfar2023phononicNJP,kianfar2023local}. These boundary conditions are valid if the PSub motion is only in the wall-normal direction and $\eta \ll \delta$ where $\eta$ is the wall-normal displacement of the PSub. Hence, throughout DNS the roughness Reynolds number is monitored and maintained below 25~\cite{morkovin1990roughness}. \\
\indent Each PSub unit is modeled as a finite linear elastic metamaterial consisting of 5 rod unit cells with one local mass-spring resonator in the center of the unit cell~\cite{kianfar2023phononicNJP,kianfar2023local}.~The PSub is free to deform at the edge interfacing with the flow (top) and is fixed at the other end (bottom).~Each individual PSub is allowed to deform independently from the adjacent rigid wall and from the motion of neighboring PSubs; its top surface deformation takes a uniform profile across the fluid-PSub interface region~\cite{kianfar2023local}. The length of the unit cell along the wall-normal direction is $L_\mathrm{UC}= 1$ $\mathrm{cm}$ (i.e., total PSub length is $5$ $\mathrm{cm}$). The resonator frequency is set to $\Omega_\mathrm{res}=2000$ $\mathrm{Hz}$ by tuning the resonator's point mass to be ten times heavier than the total mass of the unit-cell base ($m_\mathrm{res}=10\times \rho L_\mathrm{UC}$), where $\rho$ is the base material density. Hence, the stiffness of the resonator spring is $k_\mathrm{res}=m_\mathrm{res}(2\pi f_\mathrm{res})^2$. The base is composed of ABS polymer with density of $\rho=1200$ $\mathrm{kg/m^3}$ and Young's modulus of $E=3$ $\mathrm{GPa}$. Material  damping is modeled as viscous proportional damping with constants $q_1=0$ and $q_2=6\times 10^{-8}$~\cite{kianfar2023local}.
\\
\indent The Navier-Stokes equations are integrated using a time-splitting scheme~\cite{Dana91,Saiki93,Kucala14} on a staggered structured grid system. A two-node iso-parametric finite-element model is used for determining the PSub nodal axial displacements, velocities, and accelerations~\cite{HusseinJSV06} where time integration is implemented  simultaneously with the flow simulation using an implicit Newmark algorithm \cite{Newmark}. Since the equations for the fluid and the PSub are inverted separately in the coupled simulations, a conventional serial staggered scheme~\cite{Farhat_2000} is implemented to couple the two sets of time integration.~This scheme has been extensively verified, yielding excellent agreement with
the experimentally validated linear theory with a maximum deviation of 0.05\% in the predicted perturbation energy
growth~\cite{kucala2014spatial}.More details on the computational models and numerical schemes used are detailed in Refs.~\cite{kianfar2023phononicNJP,kianfar2023local}. The relative geometric dimensions of the PSubs forming each of the square and hexagonal lattices are to scale as shown in Fig.~\ref{fig:F4}, with more information available in the Supplemntal Materials document. \\

\noindent \textbf{Acknowledgments} \\
\indent This research project is currently being supported by Office of Naval Research Multidisciplinary University Research Initiative (MURI) Grant Number N0001421268. The presented work utilized the Summit supercomputer, which is supported by the National Science Foundation (awards ACI-1532235 and ACI-1532236), the University of Colorado Boulder, and Colorado State University. The Summit supercomputer is a joint effort of the University of Colorado Boulder and Colorado State University. This work also utilized the Alpine high performance computing resource at the University of Colorado Boulder. Alpine is jointly funded by the University of Colorado Boulder, the University of Colorado Anschutz, and Colorado State University and with support from NSF grants OAC-2201538 and OAC-2322260.
 
\bibliographystyle{ieeetr}
\bibliography{RefPSubs}

\end{document}


\title{SUPPLEMENTARY MATERIAL \\
\vspace{5mm}
Scatterless interference: \\
Delay of laminar-to-turbulent flow transition by a lattice of subsurface phonons}

\author{Mahmoud I. Hussein}
\email[]{mih@colorado.edu}
\affiliation{Smead Department of Aerospace Engineering Sciences, University of Colorado, Boulder, Colorado 80303}
\affiliation{Department of Physics, University of Colorado, Boulder, Colorado 80302}
\author{David Roca}
\affiliation{Centre Internacional de M\`{e}todes Num\`{e}rics en Enginyeria (CIMNE), Barcelona 08034, Spain}
\affiliation{Universitat Polit\`{e}cnica de Catalunya, ESEIAAT Campus Terrassa UPC, Terrassa 08222, Spain}
\author{Adam R. Harris}
\affiliation{Smead Department of Aerospace Engineering Sciences, University of Colorado, Boulder, Colorado 80303}
\author{Armin Kianfar}
\affiliation{Smead Department of Aerospace Engineering Sciences, University of Colorado, Boulder, Colorado 80303}

\maketitle

This Supplementary Material document covers the mathematical development and computational implementation of Bloch's theorem~\cite{Bloch1929,hussein2014dynamics} for the flow perturbation problem in the presence of a phononic subsurface (PSub). For comparison, the application of Bloch's theorem to the conventional acoustic wave propagation problem is also presented.  In the acoustics problem, acoustic pressure waves experience interferences by a lattice of elastic pillars placed above the surface, i.e., within the same spatial domain of the interfering waves (Section \ref{Acoustics}). In the flow problem, on the other hand, flow perturbation waves experience interferences by a lattice of PSub units, which, by definition, are placed beneath the surface, i.e., within a separate spatial domain to that of the interfering waves (Section \ref{FlowStability}). 

\section{Bloch analysis of acoustic waves scattered by lattice of elastic pillars: \\ \textit{Lattice placed above the surface}}
\label{Acoustics}

First we analyze the propagation of acoustic pressure waves through a lattice of elastic scatterers. This is a widely studied problem~\cite{sigalas1996attenuation}, but provides a contrast to our PSub-flow interaction problem which we cover below. Here the scatterers are in the same spatial domain as the waves being scattered, i.e., above the surface. We consider squared and hexagonal lattice configurations of elastic pillars. More specifically, the scatterers are modelled as infinite pillars in the vertical direction with a squared cross-section of size $b$ and periodically distributed with a center-to-center distance $a$. The values of $a$ and $b$, as well as the material properties of these scatterers, have been taken as those for the corresponding PSub lattice in the flow problem. The same fluid properties have been considered for the acoustic domain, i.e., the space between the scatterers. The coupled elastoacoustic problem has been solved for a 2D horizontal slice in the frequency domain using COMSOL~\cite{comsol2023}. The model ensures compatibility at the acoustic-solid interfaces by imposing
\begin{gather}
    \omega^2\mathbf{u}\cdot\mathbf{n} = \partial_n p,\\
    \mathbf{f} = -p\mathbf{n},
\end{gather}
where $\mathbf{n}$ is the outward normal vector at the solid boundaries, $\mathbf{u}$ and $\mathbf{f}$ denote the the displacement field and external force's amplitudes in the solid domain, and $p$ refers to the acoustic pressure's amplitude.

\begin{figure}
\centering
\includegraphics{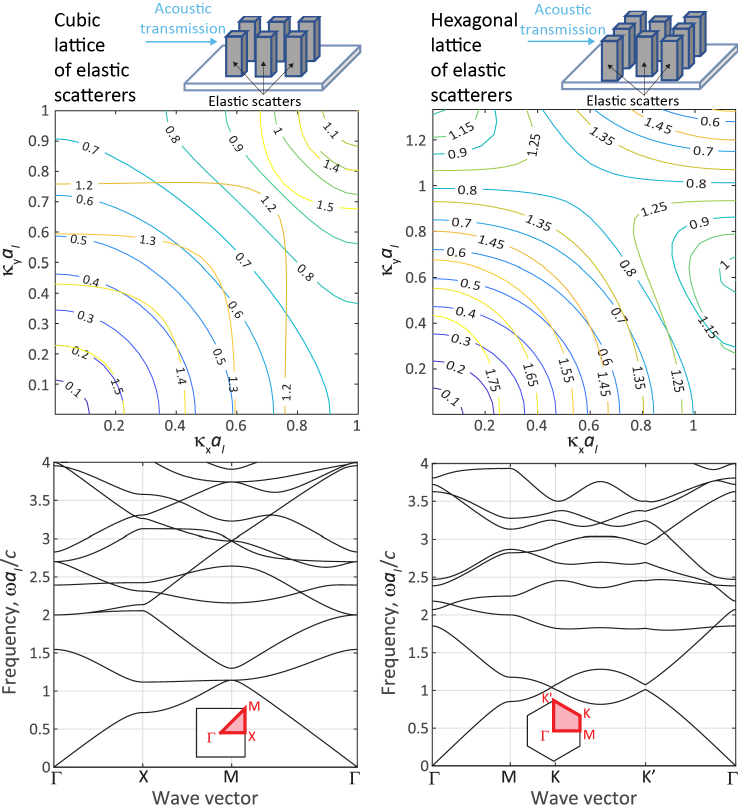}
\caption{Top row shows a schematic representation of the elastoacoustic problem of an acoustic wave travelling through a periodic lattice of elastic scatterers in the form of infinitely long pillars arranged in squared (left) and hexagonal (right) configurations. The middle and bottom plots show the isofrequency contours and acousic band structure diagrams, respectively, obtained for each case. The irreducible Brillouin zones in the reciprocal space are depicted in the insets within the bottom plots.}\label{fig:s1}
\end{figure}

To obtain the acoustic band structure characteristics, a dispersion analysis has been performed on a unit cell imposing Bloch boundary conditions at its edges, namely
\begin{equation}
    p(\mathbf{x}+\mathbf{a}) = p(\mathbf{x})e^{\text{i}\boldsymbol{\kappa}\cdot\mathbf{a}},
\end{equation}
where $\mathbf{a}$ is the periodicity vector and $\boldsymbol{\kappa}$ is the wavevector. The results are given in Fig.~\ref{fig:s1} and reveal the presence of several Bragg scattering partial band gaps when considering acoustic transmission in the streamwise $x$-direction ($\Gamma$-X portion in the squared lattice case, and $\Gamma$-M section in the hexagonal lattice case). To better appreciate the effects of these band gaps, and of the scattering mechanisms that lead to the acoustic wave's amplitude attenuation, a second analysis has been performed in a truncated configuration to three unit cells in the streamwise direction. In this study, periodic boundary conditions are still applied on the top and bottom boundaries of the acoustic domain. On the left (upstream) edge of the domain, an incident pressure plane wave traveling in the streamwise direction is imposed. A perfectly matched layer is imposed on the right edge at a far enough downstream distance from the scatterers, to model the infinite extension of the domain. The results for this analysis at a frequency lying inside the first band gap for both the squared and hexagonal lattices are found in Fig. 4 in the main article.

\section{Bloch analysis of flow perturbation waves influenced by lattice of phononic subsurface units: \\ \textit{Lattice placed beneath the surface}}
\label{FlowStability}


Now we examine our main problem, namely, undergoing Bloch analysis of the flow perturbations (instability waves) traveling in a channel where the bottom wall includes a lattice of PSubs in the subsurface. This problem is set up as a stability analysis of the Navier-Stokes equations applied to a channel flow with a given Reynolds number,
\begin{equation}
    \text{Re} = \dfrac{\rho_\text{f}U_\text{c}\delta}{\mu_\text{f}},
\end{equation}
where $\rho_\text{f}$ and $\mu_\text{f}$ are the fluid's density and viscosity, $U_\text{c}$ is the flow speed at the centre of the channel (considered the reference velocity) and $\delta$ is half the channel's height (used as reference length). A Tollmien--Schlichting (T-S) perturbation is assumed to propagate along the flow, characterized by a frequency $\omega_\text{TS}$ and a wavelength $\lambda_\text{TS} = 2\pi\delta$.   

On the bottom wall, PSub structural units are distributed along the $xz$-plane as a lattice positioned beneath the surface where each unit has its top end exposed to the flow. For the purpose of this analysis, the interfaces between the PSubs and the flow are modeled as squared patches with side length $b = \lambda_\text{TS}/4$. The PSub units are arranged in a squared or a hexagonal lattice, with a center-to-center distance $a = \lambda_\text{TS}/2$. In the forthcoming analysis, the same data used in the main article has been considered, including the Reynolds number $\text{Re}=7500$ and the parameters $\delta = 4.38\times10^{-4}$~m, $\rho_\text{f} = 1000$~kg/m${}^3$ and $\mu_\text{f} = 10^{-3}$~Pa$\cdot$s.

\subsection{Modelling the PSub response}
The PSub structure is assumed to have a one-dimensional (1D) response in the $y$-direction normal to the wall. This response is modeled in terms of an axial displacement $\eta$ driven by the governing equations of a structure comprising an elastic slender rod with attached resonators (see Fig. 2 in main article). In matrix form, the FE--discretized system of equations yields
\begin{equation}\label{eq_PSub}
    [\mathbf{M}]\{\ddot{\boldsymbol{\eta}}\} + [\mathbf{C}]\{\dot{\boldsymbol{\eta}}\} + [\mathbf{K}]\{\boldsymbol{\eta}\} = \{\mathbf{f}\},
\end{equation}
where $[\mathbf{M}]$ is the mass matrix, $[\mathbf{K}]$ is the stiffness matrix, and $[\mathbf{C}]$ is the damping matrix. The system is excited at the flow interface by the perturbation pressure from the fluid, so $f(0,t) = -p_\text{w}(t)$ (the negative sign indicates that the force acts opposite to the outward normal to the wall). On the other end of the Psub structure, the displacement is fixed $\eta(-L,t) = 0$. The sub-index `$\text{w}$' is used to denote the PSub-flow interface, i.e., $y=0$. Given the 1D nature of the system, everything is normalized by the the PSub unit/patch cross-sectional surface area.

Assuming harmonic excitation at a given frequency $\omega$,
\begin{equation}
    p_\text{w}(t) = \bar{p}_\text{w} e^{-\text{i}\omega t} + c.c.,
\end{equation}
the system in Eq.~\eqref{eq_PSub} can be solved to find the amplitude of the displacement at the flow interface, $\bar{\eta}_\text{w}$. This allows us to define the admittance of the PSub as
\begin{equation}\label{eq_Y}
    Y_\text{PSub} = \dfrac{\dot{\eta}_\text{w}}{p_\text{w}} \ \ \rightarrow \ \ \bar{Y}_\text{PSub} = \dfrac{-\text{i}\omega \bar{\eta}_\text{w}}{\bar{p}_\text{w}}.
\end{equation}
It is worth noting that $\bar{Y}_\text{PSub}$ can be obtained, for a given frequency, by solving the system in Eq.~\eqref{eq_PSub} in the frequency domain with $\bar{p}_\text{w} = 1$. From Eq.~\eqref{eq_Y}, it can be seen that the product between the amplitude and the phase of $\bar{Y}_\text{PSub}$ is proportional to the performance metric used to characterize the PSub (See Fig. 2 in main article).

\subsection{Model of a unit cell in the PSub-influenced fluid domain}
In what follows, partial derivatives will be denoted by $\partial_\phi(\bullet) = \partial(\bullet)/\partial\phi$. In this regard, the gradient operator will be defined as $\boldsymbol{\nabla} = (\partial_x, \partial_y, \partial_z)$ and the Laplacian operator $\Delta = \partial_x^2 + \partial_y^2 + \partial_z^2$. Starting with the three-dimensional (3D) Navier-Stokes equations, we write
\begin{gather}
    \boldsymbol{\nabla} \cdot \mathbf{V} = 0, \label{eq_NS1}\\
    \partial_t \mathbf{V} + (\mathbf{V}\cdot\boldsymbol{\nabla})\mathbf{V} + \boldsymbol{\nabla}P - \dfrac{1}{\text{Re}}\Delta\mathbf{V} = \mathbf{0}, \label{eq_NS2}
\end{gather}
where $\mathbf{V} = (U,V,W)$ are the fluid velocity components, and $P$ refers to the pressure field. We will assume the velocity and pressure fields can be decomposed into the base flow components $\mathbf{v}_\text{b}$ and $p_\text{b}$, and \textit{small} perturbation components $\mathbf{v}=(u,v,w)$ and $p$, such that
\begin{gather}
    \mathbf{V} = \mathbf{v}_\text{b} + \mathbf{v},\label{eq_V}\\
    P = p_\text{b} + p.\label{eq_P}
\end{gather}
The base flow components are assumed to satisfy Eq.~\eqref{eq_NS1} and Eq.~\eqref{eq_NS2} for an incompressible parallel flow in the $x$--direction (channel flow), hence $\mathbf{v}_\text{b} = (u_\text{b}(y), 0, 0)$,
with
\begin{equation}
    u_\text{b} = 1-(1-y)^2, \ \ \ \ u'_\text{b} = 2(1-y), \ \ \ \ u''_\text{b} = -2.
\end{equation}
Substituting Eq.~\eqref{eq_V} and Eq.~\eqref{eq_P} into Eq.~\eqref{eq_NS1} and Eq.~\eqref{eq_NS2} and neglecting second order terms we get
\begin{equation}\label{eq_NS1p}
    \partial_x u + \partial_y v + \partial_z w = 0,
\end{equation}
and
\begin{subequations}\label{eq_NS2p}
\begin{gather}
    \partial_t u + u_\text{b}\partial_x u + u'_\text{b}v + \partial_x p - \dfrac{1}{\text{Re}} \Delta u = 0,\label{eq_u}\\
    \partial_t v + u_\text{b}\partial_x v +  \partial_y p - \dfrac{1}{\text{Re}} \Delta v = 0,\label{eq_NS2p_v}\\
    \partial_t w + u_\text{b}\partial_x w + \partial_z p - \dfrac{1}{\text{Re}} \Delta w = 0.\label{eq_w}
\end{gather}    
\end{subequations}

To eliminate the pressure terms, first, we take the divergence of Eq.~\eqref{eq_NS2p}, which yields
\begin{equation}\label{eq_p1}
    \Delta p = -2 u'_\text{b} \partial_x v.
\end{equation}
An alternative pressure equation, which will be convenient in the forthcoming derivations, can be obtained by subtracting the partial derivative $\partial_y$ of Eq.~\eqref{eq_NS2p_v} from Eq.~\eqref{eq_p1},
\begin{equation}\label{eq_p2}
    (\partial_x^2+\partial_z^2)p = \left( \partial_t\partial_y + (u_\text{b} \partial_y - u'_\text{b})\partial_x - \dfrac{1}{\text{Re}}\partial_y \Delta \right) v.
\end{equation}
Then, by taking the Laplacian of Eq.~\eqref{eq_NS2p_v} and substituting Eq.~\eqref{eq_p1}, we get
\begin{equation}\label{eq_v}
    \left((\partial_t + u_\text{b} \partial_x)\Delta - u''_\text{b}\partial_x - \dfrac{1}{\text{Re}} \Delta^2 \right) v = 0,
\end{equation}
where $\Delta^2 = \partial_x^4 + \partial_y^4 + \partial_z^4 + 2(\partial_y^2\partial_z^2 + \partial_z^2\partial_x^2 + \partial_x^2\partial_y^2)$. It is worth noting that once $v$ is obtained from Eq.~\eqref{eq_v}, then either Eq.~\eqref{eq_p2} or Eq.~\eqref{eq_p1} can be solved to get the pressure field. Once $p$ and $v$ are known, the $u$ and $w$ components of the perturbation velocity field can be obtained from Eq.~\eqref{eq_u} and Eq.~\eqref{eq_w}, respectively.

To deal with the temporal term in the partial differential equations, the system will be expressed in the frequency domain, hence
\begin{gather}
    \mathbf{v}(\mathbf{x},t) = \bar{\mathbf{v}}(\mathbf{x}) e^{-\text{i}\omega t} + c.c., \label{eq_vt}\\
    p(\mathbf{x},t) = \bar{p}(\mathbf{x}) e^{-\text{i}\omega t} + c.c.
\end{gather}
Furthermore, assuming the perturbation propagates in the $x$-direction, 
\begin{gather}
    \bar{\mathbf{v}}(\mathbf{x}) = \tilde{\mathbf{v}}(\mathbf{x}) e^{\text{i}\kappa x},\\
    \bar{p}(\mathbf{x}) = \tilde{p}(\mathbf{x}) e^{\text{i}\kappa x},
    \label{eq_pb}
\end{gather}
where $\kappa$ refers to the wavenumber of the T-S perturbation. The combination of Eqs.~\eqref{eq_vt}-\eqref{eq_pb} renders a plane wave solution. 

Thus far, the formulation follows the standard Orr-Sommerfeld approach~\cite{Orr1907,Sommerfeld1909,Benjamin_1960}. Now we account for the periodic arrangement of the PSubs in the $xz$--plane, and apply Bloch's theorem~\cite{Bloch1929,hussein2014dynamics} to a single unit cell representing the flow domain over a single PSub with the fluid-structure interaction accounted for. It follows that the amplitude fields $\tilde{\mathbf{v}}$ and $\tilde{p}$ must be periodic in the $xz$ spatial domain. To impose the periodicity required by Bloch's theorem, a two-dimensional (2D) Fourier expansion is considered, hence
\begin{gather}
    \tilde{\mathbf{v}}(\mathbf{x}) = \sum_n \hat{\mathbf{v}}_n(y) e^{\text{i}(\alpha_n x + \beta_n z)}, \label{eq_vp}\\
    \tilde{p}(\mathbf{x}) = \sum_n \hat{p}_n(y) e^{\text{i} (\alpha_n x + \beta_n z)},
\end{gather}
where $\hat{\mathbf{v}}_n$ and $\hat{p}_n$ are the amplitude coefficients associated with the $n$-th component in the series, and $\alpha_n$ and $\beta_n$ are the corresponding Fourier wavenumbers. In general, for a truncated series with $2N_x+1$ and $2N_z+1$ terms in each direction, each $n$-th component in the summation will be associated with a pair of indices $(i,j)_n$ in the ranges $i = \{-N_x, ... , 0, ... , N_x\}$ and $j = \{-N_z, ... , 0, ... , N_z\}$. In this regard, we have
\begin{equation}
    \alpha_n = \left\{
        \begin{array}{ll}
            i_n \dfrac{2\pi}{a}, & \text{for the squared lattice,} \\
            \dfrac{2i_n-j_n}{\sqrt{3}} \dfrac{2\pi}{a}, & \text{for the hexagonal lattice,}
        \end{array}
    \right. \ \text{and} \ \ 
    \beta_n = j_n \dfrac{2\pi}{a}.
\end{equation}

With these definitions, we can deal with the partial derivatives $\partial_t \equiv -\text{i}\omega$, $\partial_x \equiv \text{i}(\kappa + \alpha_n)$ and $\partial_z \equiv \text{i}\beta_n$. Then, substituting into Eq.~\eqref{eq_v}, we get a fourth-order ordinary differential equation for each $n$ component in the series expansion, in terms of the Fourier amplitude $\hat{v}_n$ (which is only a function of the spatial $y$--coordinate),
\begin{equation}\label{eq_OS}
    \left((\omega - u_\text{b}(\kappa + \alpha_n))\Delta + u''_\text{b}(\kappa + \alpha_n) + \dfrac{\text{i}}{\text{Re}}\Delta^2\right)\hat{v}_n = 0,
\end{equation}
where $\Delta = \partial_y^2 - ((\kappa + \alpha_n)^2 + \beta_n^2)$ and $\Delta^2 = \partial_y^4 -2((\kappa + \alpha_n)^2+\beta_n^2) \partial_y^2 + ((\kappa + \alpha_n)^2 + \beta_n^2)^2$. It is worth noting that Eq.~\eqref{eq_OS} is the generalized Orr-Sommerfeld equation to a 3D spatial domain. In fact, one recovers the classical 1D version by making $\alpha_n = \beta_n = 0$.
 
To account for the PSub interaction with the flow, no--slip transpiration boundary conditions are applied~\cite{kianfar2023phononicNJP}. For the case of a vibrating wall with small displacements in the normal $y$--direction, these take the form of
\begin{subequations}\label{eq_bcc}
    \begin{align}
        u(x,0,z,t) &= -u'_\text{b}(0)\eta_\text{w} \mathcal{H}(x,z),\\
        v(x,0,z,t) &= \dot{\eta}_\text{w} \mathcal{H}(x,z),\\
        w(x,0,z,t) &= 0,
    \end{align}
\end{subequations}
where $\eta_\text{w}$ refers to the PSub displacement, and $\mathcal{H}$ is a periodic step function defined such that $\mathcal{H} = 1$ for $(x,z) \in \Gamma_\text{PSub}$ (i.e., in the area occupied by the patches representing the top of the PSub units) and $\mathcal{H} = 0$ otherwise. Expressing Eq.~\eqref{eq_bcc} in the frequency domain and using the definition in Eq.~\eqref{eq_Y}, we get
\begin{subequations}\label{eq_bcc_f}
    \begin{align}
        \text{i} \omega \Tilde{u}(x,0,z) &= u'_\text{b}(0)  \bar{Y}_\text{PSub} \Tilde{p}_\text{w} \mathcal{H}(x,z),\\
        \Tilde{v}(x,0,z) &= \bar{Y}_\text{PSub} \Tilde{p}_\text{w} \mathcal{H}(x,z),\\
        \Tilde{w}(x,0,z) &= 0.
    \end{align}
\end{subequations}
It is worth noting that the T-S wave spatial perturbation component has also been accounted for in Eq.~\eqref{eq_bcc_f}, assuming $\bar{p}_\text{w} = \Tilde{p}_\text{w} e^{\text{i}\kappa x}$. To proceed, the 2D Fourier expansion is applied, yielding
\begin{subequations}\label{eq_bcc_n}
    \begin{align}
        \text{i} \omega \hat{u}_n(0) &= u'_\text{b}(0)  \bar{Y}_\text{PSub} \Tilde{p}_\text{w} \hat{\mathcal{H}}_n, \label{eq_bcc_u}\\
        \hat{v}_n(0) &= \bar{Y}_\text{PSub} \Tilde{p}_\text{w} \hat{\mathcal{H}}_n, \label{eq_bcc_v}\\
        \hat{w}_n(0) &= 0,
    \end{align}
\end{subequations}
where $\hat{\mathcal{H}}_n$ are the Fourier coefficients for the step function, given by
\begin{equation}
    \hat{\mathcal{H}}_n = \dfrac{1}{a^2} \int_{-b/2}^{b/2} \int_{-b/2}^{b/2} e^{\text{i}(\alpha_n x + \beta_n z)} \text{d}x\text{d}z = \dfrac{\sin{(\alpha_n b/2)}}{\alpha_n a/2} \dfrac{\sin{(\beta_n b/2)}}{\beta_n a/2}.
\end{equation}
Using Eq.~\eqref{eq_NS1p}, the boundary condition Eq.~\eqref{eq_bcc_u} can be re-written as
\begin{equation}
    \omega \partial_y \hat{v}_n(0) = -(\kappa + \alpha_n) u'_\text{b}(0) \hat{v}_n(0).
\end{equation}
To make the analysis consistent with the FSI simulations, the pressure amplitude value that triggers the PSub response is taken as the average over $\Gamma_\text{PSub}$, hence
\begin{equation}\label{eq_bcc_1}
    \Tilde{p}_\text{w} = \dfrac{1}{b^2} \int_{-b/2}^{b/2} \int_{-b/2}^{b/2} \Tilde{p}(x,0,z) \text{d}x\text{d}z = \dfrac{a^2}{b^2} \sum_m \hat{p}_m(0) \hat{\mathcal{H}}_m.
\end{equation}
Equation Eq.~\eqref{eq_p2} can be used to express the pressure coefficients at the wall in terms of the vertical component of the perturbation velocity, such that
\begin{equation}
    \hat{p}_m(0) = \dfrac{1}{((\kappa + \alpha_m)^2+\beta_m^2)\text{Re}}\left( \partial_y^3 - ((\kappa + \alpha_m)^2+\beta_m^2)\partial_y  \right) \hat{v}_m(0).
\end{equation}
Then, the boundary condition Eq.~\eqref{eq_bcc_v} can be expressed as
\begin{equation}\label{eq_bcc_2}
    \hat{v}_n(0) = \dfrac{\bar{Y}_\text{PSub}}{\text{Re}} \dfrac{a^2}{b^2} \sum_m \left( \dfrac{\hat{\mathcal{H}}_m \hat{\mathcal{H}}_n}{(\kappa + \alpha_m)^2+\beta_m^2}\left( \partial_y^3 - ((\kappa + \alpha_m)^2+\beta_m^2)\partial_y  \right) \hat{v}_m(0) \right).
\end{equation}
The introduction of the PSub admittance as a boundary condition has been used in Ref.~\cite{Barnes_2021} in the context of a similar stability analysis but considering only a lone PSub occupying the entire boundary layer, as opposed to a lattice of PSubs. 

Noticeably, by setting $\bar{Y}_\text{PSub} = 0$, the conventional rigid wall boundary conditions are recovered. These are applied on the top wall as
\begin{align}
    \hat{v}_n(2) &= 0, \label{eq_bcc_1t}\\
    \partial_y \hat{v}_n(2) &= 0\label{eq_bcc_2t}.
\end{align}

The whole set of Eq.~\eqref{eq_OS}, with the corresponding boundary conditions provided by Eq.~\eqref{eq_bcc_1} and Eq.~\eqref{eq_bcc_2} for the bottom wall and by Eq.~\eqref{eq_bcc_1t} and Eq.~\eqref{eq_bcc_2t} for the top wall, represent a system that can be solved for a given frequency $\omega$ and the associated PSub admittance $\bar{Y}_\text{PSub}$ as a generalized eigenvalue problem, in which the wavenumber $\kappa$ becomes the eigenvalue and each $\hat{v}_n$ represents an eigenfunction. It is worth noting that in order for $\kappa$ to represent the eigenvalue, first Eq.~\eqref{eq_bcc_2} must be multiplied by the common denominator, which increases the order of the generalized eigenvalue problem. Regardless, a solution can be found numerically upon applying some form of spatial discretization of the terms $\hat{v}_n$. In this work, an FE based scheme with standard Hermite elements$-$to capture the higher-order derivatives involved$-$is considered.

For each eigenvalue $\kappa$, one can obtain from the associated eigenfunctions $\hat{v}_n$ (or eigenvectors in the discretized version of the problem) the corresponding pressure coefficients $\hat{p}_n$ from Eq.~\eqref{eq_p1} or Eq.~\eqref{eq_p2}. Then, the remaining perturbation velocity components $\hat{u}_n$ and $\hat{w}_n$ can be obtained from Eq.~\eqref{eq_u} and Eq.~\eqref{eq_w}, respectively. Alternatively, Eq.~\eqref{eq_NS1p} can be used once either $\hat{u}_n$ or $\hat{w}_n$ has been obtained to find the other. 

\begin{figure} [b!]
\centering
\includegraphics[width=0.825\textwidth]{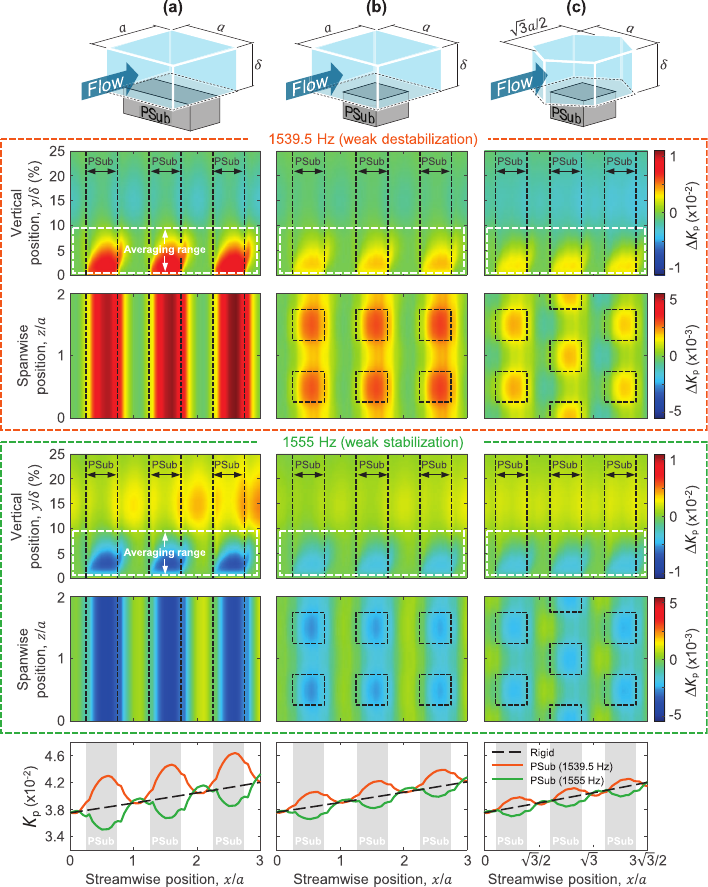}
\caption{Perturbation kinetic energy distributions corresponding to the unstable mode for the (a) full-span, (b) square lattice, and (c) hexagonal lattice configurations depicted in the top row. Results for each frequency are depicted in their corresponding panel. The first panel (in orange) shows a weak destabilizing case (at a frequency of 1539.5~Hz) and the second panel (in light green) is for a weak stabilizing case (at a frequency of 1555~Hz). In each panel, the top row depicts the average (over the entire spanwise direction) vertical distribution of the modal perturbation KE difference between the PSub case and the rigid case, denoted $\Delta K_{\text{p}}$, while the bottom row shows the horizontal distribution of the same property, averaged in the vicinity of the PSub wall (where the effect is stronger), up to 10\% of half of the channel's height (see region highlighted in white in the vertical distribution plots). The bottom plots below the panels summarize the averaged modal perturbation KE evolution obtained for both the weak destabilizing and stabilizing cases compared to that of the rigid case. As a reference, all the modal amplitudes have been normalized to make them coincide with that of the rigid case at the left edge of the first unit cell.}\label{fig:s2}
\end{figure}

\begin{figure} [b!]
\centering
\includegraphics[width=1\textwidth]{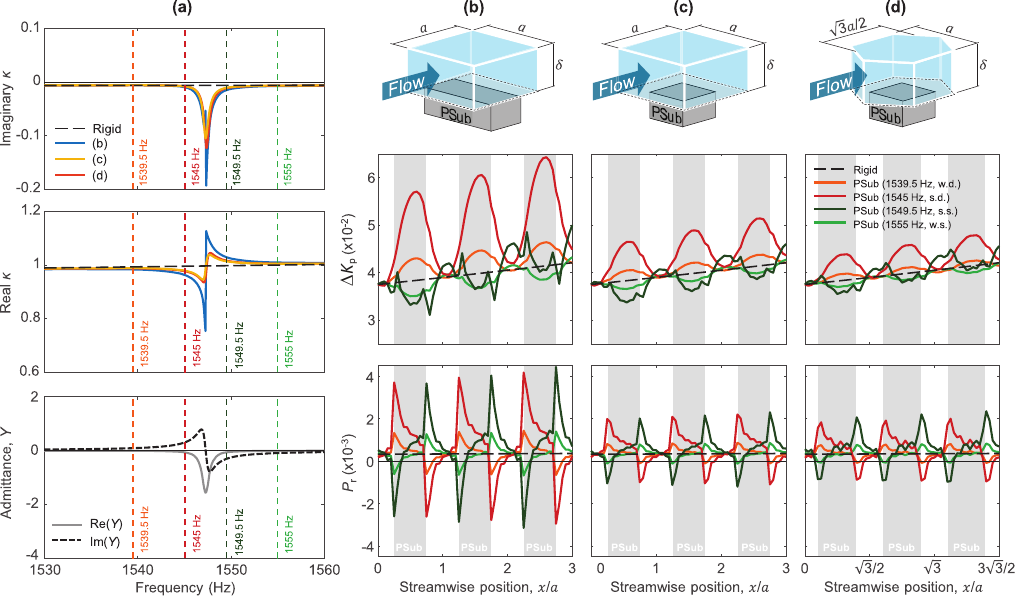}
\caption{(a) Dispersion curves corresponding to the unstable mode. The first and second rows correspond to the imaginary and real components of the wavenumber $\kappa$ for the rigid cases and the PSub cases in the full-span, square lattice, and hexagonal lattice configurations [depicted by the (b), (c) and (d) unit cell schematics, respectively]. The third row shows the real and imaginary components of the PSub's admittance $Y$ at different frequencies. The dashed vertical lines indicate the frequencies for which the average modal perturbation KE and their corresponding rates of production $P_\text{r}$ curves on the right plots have been obtained. In particular, weak and strong destabilizing (in orange and dark red, respectively) and weak and strong stabilizing (in light and dark green, respectively) have been considered.~The perturbation KE and production rate curves correspond to averages over the entire spanwise direction and the first 10\% section of half of the channel's height in the vicinity of the PSub wall (as in Fig.~\ref{fig:s2}). The modal amplitudes have been normalized to make them coincide with that of the rigid case at the left edge of the first unit cell in each case.}\label{fig:s3}
\end{figure}

We show the results of our flow perturbation Bloch analysis in Figs.~\ref{fig:s2} and~\ref{fig:s3} and Fig. 3 in the main article. Bloch's theorem has been widely used for acoustic and elastic wave propagation problems, as demonstrated in Section~\ref{Acoustics}. The theorem has recently also been applied to a flow stability problem involving a repeated 1D array of rigid riblets in a channel~\cite{jouin2024modal}. In that work, the analysis domain comprised several repeated cells, and the focus has been on calculating the complex frequency response for a given wavenumber. Here, we examine a 2D lattice of PSubs placed in the subsurface, with the fluid-structure interaction accounted for, and considering both square and hexagonal symmetries. We also limit our analysis to a single unit cell, as commonly done in Bloch analysis. Furthermore, we calculate the flow perturbations dispersion curves for the most unstable mode. In the main article, we discuss, and demonstrate, the benefit of this analysis on guiding the PSub lattice design to achieve delay of flow transition. 
\subsection{Energetics of the Bloch solution}
To focus the Bloch flow analysis on the most relevant mode, we select the closest to the unstable mode resulting from the classical Orr-Sommerfeld equation, which has been used as input of the T-S wave in the FSI simulations. In our framework, this can be obtained simply by setting $\bar{Y}_\text{PSub} = 0$ on the bottom wall boundary conditions. Tracking this mode for a selected range of frequencies allows us to obtain the stability plots given in Figure~\ref{fig:s3}(a). The results show a correlation between the real and imaginary components of the admittance with the imaginary and real components of the unstable wavenumber, respectively. As the value of the admittance increases, the PSub effects are more noticeable, consistently with what the PSub performance metric indicates. In order to visualize the destabilizing and stabilizing effects, we evaluate for the associated mode the perturbation kinetic energy (KE) $K_{\text{p}}$, as well as the rate of production of this quantity, which we denote as $P_\text{r}$. To do so, once the corresponding $\hat{v}_n$ and $\hat{u}_n$ terms have been obtained, Eq.~\eqref{eq_vp} is used to get the complex amplitude of the mode at a given point, which can be expressed in terms of its absolute value $|\Tilde{\mathbf{v}}(\mathbf{x})|$ and phase $\varphi_{\Tilde{\mathbf{v}}}(\mathbf{x})$ as
\begin{subequations}\label{eq_bcc_n}
    \begin{align}
        \Tilde{u}(\mathbf{x}) &= |\Tilde{u}(\mathbf{x})| e^{\text{i}\varphi_{\Tilde{u}}(\mathbf{x})},\\
        \Tilde{v}(\mathbf{x}) &= |\Tilde{v}(\mathbf{x})| e^{\text{i}\varphi_{\Tilde{v}}(\mathbf{x})},\\
        \Tilde{w}(\mathbf{x}) &= |\Tilde{w}(\mathbf{x})| e^{\text{i}\varphi_{\Tilde{w}}(\mathbf{x})}.
    \end{align}
\end{subequations}
The T-S wave propagation can then be defined using Eq.~\eqref{eq_vt} as
\begin{subequations}\label{eq_bcc_n}
    \begin{align}
        u(\mathbf{x},t) &= 2|\Tilde{u}(\mathbf{x})| e^{-\kappa_\text{I} x} \cos{(\kappa_\text{R}x - \omega t + \varphi_{\Tilde{u}}(\mathbf{x}))},\\
        v(\mathbf{x},t) &= 2|\Tilde{v}(\mathbf{x})| e^{-\kappa_\text{I} x} \cos{(\kappa_\text{R}x - \omega t + \varphi_{\Tilde{v}}(\mathbf{x}))},\\
        w(\mathbf{x},t) &= 2|\Tilde{w}(\mathbf{x})| e^{-\kappa_\text{I} x} \cos{(\kappa_\text{R}x - \omega t + \varphi_{\Tilde{w}}(\mathbf{x}))},
    \end{align}
\end{subequations}
where $\kappa=\kappa_\text{R} + \text{i}\kappa_\text{I}$ has been considered. Then, the average position-dependent modal perturbation KE over one period of time $T=2\pi/\omega$ is obtained as
\begin{equation}
    K_{\text{p}}(\mathbf{x}) = \dfrac{1}{T} \int_0^T \dfrac{1}{2} \left( \left[ u(\mathbf{x},t) \right]^2 + \left[ v(\mathbf{x},t) \right]^2 + \left[ w(\mathbf{x},t) \right]^2 \right) \text{d}t = \left( |\Tilde{u}(\mathbf{x})|^2 + |\Tilde{v}(\mathbf{x})|^2 + |\Tilde{w}(\mathbf{x})|^2 \right) e^{-2 \kappa_\text{I} x}.
\end{equation}
In a similar fashion, the average position-dependent rate of production of modal perturbation KE is computed as
\begin{equation}\label{eq_pr}
    P_\text{r}(\mathbf{x}) = -\dfrac{1}{T} \int_0^T u(\mathbf{x},t)v(\mathbf{x},t)u'_\text{b} \text{d}t = -2|\Tilde{u}(\mathbf{x})||\Tilde{v}(\mathbf{x})|u'_\text{b} e^{-2\kappa_\text{I} x} \cos{(\varphi_{\Tilde{u}}(\mathbf{x}) - \varphi_{\Tilde{v}}(\mathbf{x}))}.
\end{equation}
From equation Eq.~\eqref{eq_pr}, it can be seen that the sign of the production rate is determined by the relative phase between the horizontal and vertical components of the perturbation velocity, being positive when they are in-phase, and negative when they are out-of-phase. 

To show the consistency of this analysis with the predicted effects of the PSub by the performance metric of Fig. 2 in the main article, Fig.~\ref{fig:s2} shows the computed $K_{\text{p}}$ obtained from the perturbation velocity components of the unstable mode for the different PSub configurations (full-span, square lattice, and hexagonal lattice). Two frequencies have been selected for the analysis, one at 1539.5~Hz---corresponding to a destabilizing case$-$and another at 1555~Hz$-$corresponding to a stabilizing case. To better visualize the effects, the modal amplitudes in each case have been normalized so that the average perturbation KE at the left edge of the first unit cell is the same as in the rigid case. Using the latter as a reference, the results in Fig.~\ref{fig:s2} clearly show how in the destabilizing case the modal perturbation KE increases close to the wall in the sections where the flow interacts with the PSub, with the effects being stronger in the full-span case compared to the lattice configurations. Conversely, for the stabilizing frequency, the modal perturbation KE is reduced instead.

These results are also consistent with the stability plots shown in Fig.~\ref{fig:s3}(a) where the same data are compared with those obtained at frequencies with stronger PSub interaction effects, namely at 1545~Hz and 1549.4~Hz for the strong destabilization and stabilization cases, respectively. From Fig.~\ref{fig:s3}, it can be seen that the absolute value of the admittance is higher at these frequencies, which translates into larger real and imaginary wavenumbers corresponding to the unstable mode. Consistently, both the modal perturbation KE and the associated production rates exhibit much higher amplitudes for these cases [see Fig.~\ref{fig:s3}(b)-(d)]. One can also observe how, for the same PSub admittance, the full-span configuration yields a much stronger response than the lattice cases, both in terms of the modal amplitudes of the perturbation KE and the associated imaginary wavenumber component $\kappa_\text{I}$. This can be attributed to interaction effects in the spanwise direction which open a pathway for altering the flow instabilities downstream, as demonstrated in the full-scale FSI simulations of a truncated finite set of PSub lattices. 






\bibliography{RefPSubsSupp}